# 3.5 kW coherently-combined ultrafast fiber laser

Michael Müller,[1,*] Arno Klenke,[1,2] Albrecht Steinkopff,[1,3] Henning Stark,[1] Andreas Tünnermann,[1,2,3] Jens Limpert[1,2,3]

[1]Friedrich-Schiller-University Jena, Abbe Center of Photonics, Institute of Applied Physics, Albert-Einstein-Straße 15, 07745 Jena, Germany
[2]Helmholtz-Institute Jena, Fröbelstieg 3, 07743 Jena, Germany
[3]Fraunhofer Institute for Applied Optics and Precision Engineering, Albert-Einstein-Straße 7, 07745 Jena, Germany
*Corresponding author: michael.mm.mueller@uni-jena.de

**An ultrafast laser based on coherent beam combination of four ytterbium-doped step-index fiber amplifiers is presented. The system delivers an average power of 3.5 kW and a pulse duration of 430 fs at 80 MHz repetition rate. The beam quality is excellent ($M^2$<1.24·1.10) and the relative intensity noise is as low as 1% in the frequency span from 1 Hz to 1 MHz. The system is turn-key operable as it features an automated spatial and temporal alignment of the interferometric amplification channels.**

http://dx.doi.org/10.1364/OL.43.006037

Industrial and scientific applications demand for ultrafast lasers with high peak power and high average power. This demands are met by the advanced thin-disk [1], slab [2], and fiber [3] technologies. These allow for kW-class average power due to favorable surface-to-volume ratios and GW-level peak power by using large beam diameters and chirped-pulse amplification. Nevertheless, waste heat from the laser process poses an average power limitation inducing beam quality degradation in thin-disk and slab [4] lasers and transverse mode instabilities in fiber lasers [5]. Similarly, the peak power remains limited by nonlinear effects like self-focusing, self-phase modulation and Raman scattering. Thus, the performance of an individual laser system is always clamped to a certain maximum. This limit can be surpassed by coherent beam combination of an array of laser amplifiers [6,7] in a multi-channel interferometer, as schematically depicted in Fig. 1. Instead of a single amplifier, a seed beam is divided into $N$ amplifier channels (e.g. $N$=4). The beams from these channels are combined interferometrically into a single beam. The setup has to be stabilized against environmental perturbations using phase actuators. A figure of merit is the combination efficiency, defined as the ratio of the combined beam power to all power emitted from the amplifiers. For maximum combining efficiency, the group delay, the beam parameter, and the alignment of all channels has to be matched. [8]

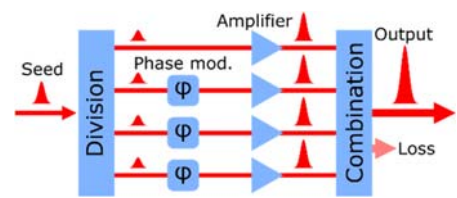

Fig. 1. Schematic of a coherently combined laser system.

Implementations of coherent beam combining can be split into tiled-aperture and filled-aperture approaches. In the first case, a large aperture beam is synthesized from a bundle of parallel beams without the need for an actual combining element [9], but with a maximum combining efficiency of 68% for Gaussian beams. In the second case, laser beams are superposed collinearly on optical elements, which can be intensity beam splitters, polarization beam splitters [10], diffractive optical elements [11], or others. These configurations theoretically allow for up to 100% combining efficiency. From these elements, particularly well high-power-suitable are partially-reflective dielectric mirrors as they are low-absorbing and easily scalable in aperture size.

Following this approach, the herein presented laser system is developed, including an automated optimization of group delay and alignment. The laser output characterization is presented and further power scaling is discussed. Potential applications of this laser system include materials processing as well as high-flux high harmonic generation after nonlinear compression [12].

The setup of the laser system is schematically depicted in Fig. 2. The seed source is an Yb-doped bulk oscillator emitting ultrashort pulses at 1040 nm central wavelength with 80 MHz repetition rate.

The seed is coupled into a polarization-maintaining fiber front-end. Here, the pulses are stretched to about 2 ns FWHM duration in chirped fiber Bragg gratings with a 15 nm spectral hard cut centered at 1047 nm. The stretched pulses pass through a pulse shaper and an adjustable pre-compressor allowing for control of spectral amplitude and phase, facilitating the pulse compression in the end [13]. Four preamplifiers are embedded in this section. The seed beam is free-space coupled into a first rod-type fiber amplifier (80 cm length, 50 μm mode-field diameter) increasing the average power to 2 W and in a second rod-type fiber amplifier (1.1 m length, 65 μm mode-field diameter) the average power is raised to 60 W. All amplifiers are optically isolated.

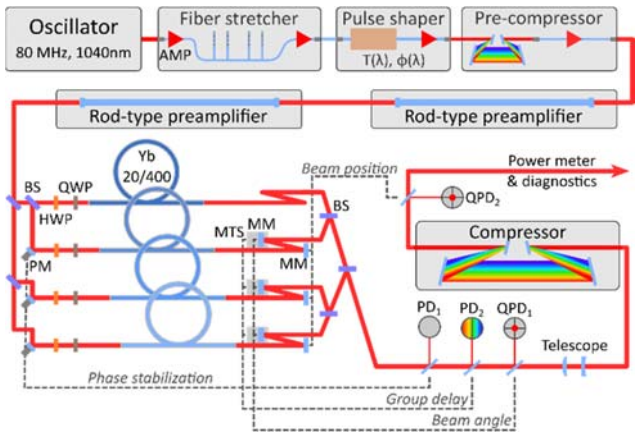

Fig. 2. Setup of the four-channel ultrafast laser system. AMP: all-fiber preamplifier, BS: beam splitter, H/QWP: half/quarter wave plate, PM: piezo-driven mirror, MM: motorized mirror, MTS: motorized translation stage, (Q)PD: (quadrant) photo diode.

The high-power seed beam is split up into four channels by means of intensity beam splitters. In three channels there is a piezo-driven mirror for phase control and in all channels there are wave plates for polarization adjustment. The beams are coupled into the main amplifiers, made from ytterbium-doped non-polarization-maintaining step-index fiber with 20 μm core diameter, 400 μm cladding diameter in a polymer coating. Each active fiber is 11 m long and coiled to 12 cm diameter to suppress higher order modes and transverse mode instabilities, is equipped with antireflective coated and wedged end caps, and is held in a water-cooled enclosure. Each amplifier is counter-pumped with up to 1.6 kW power at 976 nm (free-space coupled). After amplification, linearly polarized beams are superposed in a tree-type configuration using 50/50 beam splitters close to normal incidence. In three channels, there are each two motorized steering mirrors and a motorized translation stage to remotely adjust the beam overlap and the optical path length with respect to the non-motorized reference channel. The combined beam is steered using mirrors in water-cooled mounts to minimize thermal drift of the alignment. The weakly divergent beam is re-collimated at 6 mm $1/e^2$-diameter using a Keplerian telescope with lenses of +200 mm and -200 mm focal length. Finally, the beam passes through a Treacy-type compressor consisting of dielectric gratings with 1740 lines per mm and a transmission of 89.5%.

The implemented active phase stabilization algorithm is locking of optical coherence by single-detector electronic-frequency tagging (LOCSET, [14]). In this technique, a sinusoidal phase dither is applied to one channel of the interferometer via the respective piezo-driven mirror, creating an intensity dither at the interferometer output. The intensity dither is detected using a photo diode ($PD_1$) and its electric signal is demodulated with the initial sinusoid yielding an asymmetric error signal with zero crossings at the interference extrema of the carrier wave. This error signal is used for a feedback loop acting upon the piezo-driven mirror in that particular channel. Several interferometer channels are addressed by choosing distinct modulation frequencies, which in this experiment are 4.5 kHz, 6 kHz, and 8 kHz. The frequencies are chosen to be below the resonance frequency of the piezo mirrors (~10 kHz), but still being sufficiently separated to avoid crosstalk, and high enough to cover most phase fluctuations. Atmospheric perturbations are reduced by enclosing the setup in a housing and by dumping the combining losses at distance.

The laser system incorporates an automated optimization of spatial [15] and temporal [16] overlap, based on the already implemented phase stabilization. Information about misalignment is retrieved by LOCSET error signal detection from spatially and spectrally filtered beam samples. In other words, the spatial and spectral dither of the zeroth interference fringe is evaluated. In particular, group delay information is retrieved from a spectrally filtered sample of the output beam ($PD_2$). The spectral filter used in this experiment is an edge pass blocking half of the spectrum. Beam position and pointing misalignment is detected from horizontally and vertically filtered sections of the beam profile. Here, the spatial filtering is implemented using one quadrant photo diode ($QPD_1$) directly after the beam combination for pointing misalignment detection and another one ($QPD_2$) after the compressor for position misalignment detection. All the photo diode signals are read out and evaluated by a microcontroller that forwards the result to a computer, which in turn is used to control the motorized mirrors and the motorized translation stages. A basic optimization routine is implemented that repeatedly reads all error signals and then moves the respective components by a fixed value, with the information flow as depicted in Fig. 2. After several iterations, the alignment converges at maximum combination efficiency, which cannot be further improved on by manual adjustment.

The system is operated at several power levels, at which the pump powers are adjusted such that the amplifiers emit approximately the same average power. Changes to the output polarization are compensated for by setting the wave plates accordingly. The phase stabilization is turned on and the optimization routine is started to optimize the alignment. With each power increase, the fibers are expanding leading to significant relative path length changes, which are compensated for by the alignment mechanism. The experimental results are summarized in Table 1.

**Table 1:** Output power after compression for a single amplifier channel $P_{Ch}$ and the combined beam $P_{comb}$, the combining efficiency $\eta_{comb}$, and the lower bound nonlinear phase $B_{min}$.

| $P_{Ch}$ / W | $P_{comb}$ / W | $\eta_{comb}$ / % | $B_{min}$ / rad |
|---|---|---|---|
| 195±2 | 770 | 98.7 | 3 |
| 554±9 | 2020 | 91.2 | 7 |
| 769±18 | 2780 | 90.4 | 9 |
| 992±65 | 3500 | 88.2 | 11 |

At low output power, a very high combining efficiency of 98.7% is observed indicating a well aligned system. For increasing power, the combining efficiency decreases slowly to 88.2% and an average power of 3.5 kW after compression is achieved. The efficiency decrease can be due to increased higher order mode content of the fibers at high power or due to the very high accumulated nonlinear phase, which as a lower bound estimate is 11 rad assuming exponential amplification. At this high nonlinearity, even smallest power differences between the amplifier channels lead to significant phase distortions relative to each other.

The output spectra for the combined beam and exemplary for the single amplifier channel 1, both at highest output power, are depicted in Fig. 3. The spectrum of the single channel (yellow) features a Raman peak at 1085 nm, being the power limitation in this experiment. Strong spectral modulation of the signal in the hard cut (1038 nm to 1054 nm) is evidence for the high nonlinearity in the amplifier. In the combined beam (red), amplified spontaneous emission and Raman are suppressed as the signal power increases in each combination step whilst the incoherent part of the spectrum is not. Finally, only the signal in the compressor hard cut remains (blue).

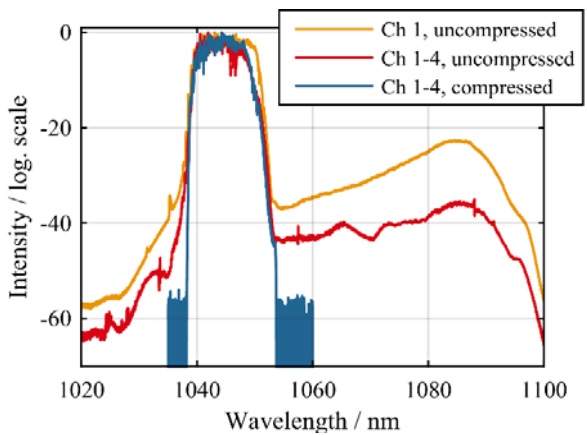

Fig. 3. Normalized optical spectra at maximum output power from amplifier channel 1 before compression and for the combined output of all four channels before and after compression.

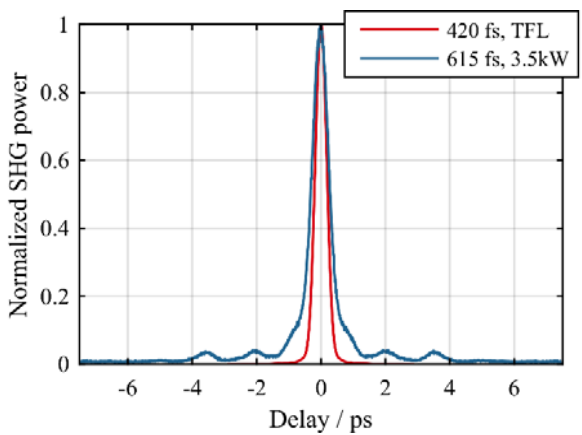

Fig. 4. Measured, background-free autocorrelation trace and calculated, transform-limited autocorrelation trace at 3.5 kW average power. The incomplete compression is due to the high amplifier nonlinearity.

The autocorrelation of the compressed pulse is depicted in Fig. 4. The FWHM duration of the autocorrelation is 615 fs, which corresponds to a 430 fs pulse duration assuming a Gaussian pulse. The transform limited pulse duration supported by the spectrum from Fig. 3 is 300 fs. The mismatch between the measured and transform limited pulse duration, despite phase shaping, is explained by the large accumulated nonlinear phase. Indeed, at low output power, the transform limited pulse duration was obtained.

Next, the beam quality of the combined beam was measured before and after pulse compression, as shown in Fig. 5 and Fig. 6. The combined beam features a diffraction limited beam quality with an $M^2$-value of less than 1.1 on both beam axes. In the compressor, a slight, but power independent beam deterioration is observed, which is traced back to an insufficient surface quality of the beam steering mirrors, stating a mere technical issue. Still, the output beam quality with $M^2$-values 1.24 and 1.10 remains very close to the diffraction limit for all power levels.

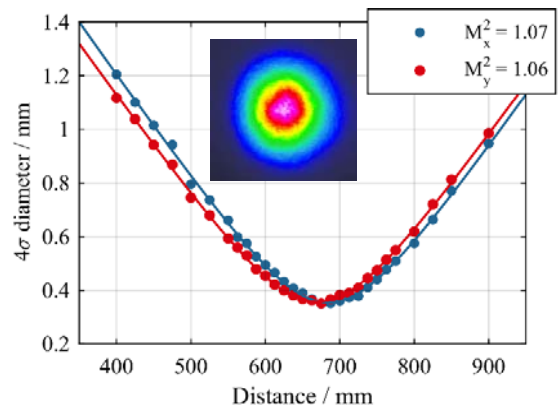

Fig. 5. $M^2$-measurement of the combined beam before compression at maximum power. Inset: Collimated beam profile before compressor.

Finally, the output power stability was analyzed using an amplified photo diode (Thorlabs PDA10CS2, 20dB gain), a 5 MHz low-pass filter, and a high resolution oscilloscope. Data was acquired with 5 MS/s for 20 s of acquisition time and processed via fast Fourier transform using a Hanning window (Equivalent noise bandwidth 1.5 Hz). The retrieved power spectral density and the integrated power spectral density are depicted in Fig. 7(a) and (b), respectively. Three cases are analyzed for operation at maximum power. In the first case (blue), the relative intensity noise of a single amplifier channel is measured. It is 0.1% in the full frequency span, which is a typical value for a fundamental mode high-power fiber amplifier. In the second case (red), two channels are interfering without stabilization. Thus, phase noise between the two channels directly transfers to intensity noise, which allows to estimate the required loop bandwidth. It is less than 1 kHz from the corresponding integrated PSD graph. (Please note, that the red integrated PSD curve is divided by 50 to be presented in the same plot.). The peak at 20 kHz arises from phase dithering that was applied for testing purpose, but ultimately has not been used in the combing experiment. The third case is the phase-stabilized combination of all four channels at maximum power (yellow). Here, the relative intensity noise amounts to 1% in the full frequency span. The excess noise originates partially from uncompensated slow phase fluctuations below 1 kHz, but mostly from the LOCSET phase dithering, its sidebands and harmonics between 4 kHz and 20 kHz. A probable reason for the low-frequency excess noise is that weak but existent interferometer phase noise in the dithering frequency range erroneously transfers into the loop. In this case, the choice of much higher dithering frequencies would be the solution, which in principle is possible, but was not technically feasible with the piezo-driven mirrors and control electronics utilized in this experiment.

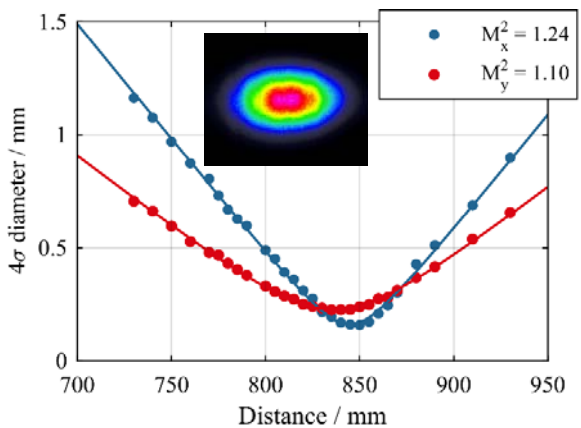

Fig. 6. $M^2$-measurement of the combined beam after compression at 3.5 kW average power. Astigmatism and beam quality deterioration are due to surface imperfections of the steering mirrors. Inset: Collimated beam profile after compressor.

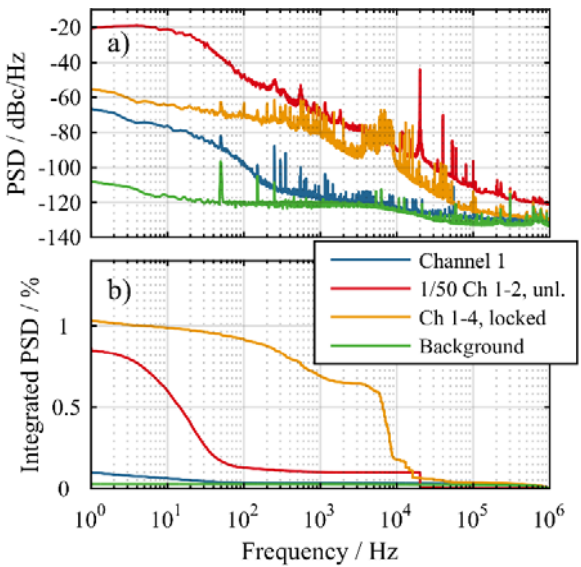

Fig. 7. Power spectral density of the output laser noise (a) and integrated power spectral density (b) for a single channel, for unstabilized superposition of two channels, for stabilized combination of all four channels and for the measurement background. The integrated PSD of the unstabilized combination of two channels has been divided by 50.

In summary, we presented an ultrafast fiber laser system based on coherent beam combination of four ytterbium-doped step-index fiber amplifier channels. The system delivers 3.5 kW average power after compression at a combining efficiency of 88.2%. The output pulse duration is 430 fs and the beam quality is excellent with $M^2$-values of 1.24 and 1.10 on the beam principal axis. The relative intensity noise is 1% in the frequency span from 1 Hz to 1 MHz. The system features automated alignment of beam position, beam pointing, and path length of the interferometer, which ensures maximum combining efficiency and allows for turn-key operation. The power limitation for a single amplifier fiber in this experiment is Raman scattering, which could be overcome by advanced large-mode-area fiber designs as well as a longer stretched pulse duration. The beam combination itself is not yet limited, as the beam splitters showed no significant heating during the experiment. Hence, power scaling by using more amplifier channels is possible right away. At the day of publication, this system features the largest average power ever reported from an ultrafast laser.

**Acknowledgment.** We acknowledge funding by the Fraunhofer Cluster of Excellence 'Advanced photon sources' and by the European Research Council Grant 670557 'MIMAS'. We like to thank the Fraunhofer IOF in Jena, Germany, for help in preparation fibers. M.M. acknowledges financial support by the Carl-Zeiss-Stiftung.

**References**
1. T. Nubbemeyer, M. Kaumanns, M. Ueffing, M. Gorjan, A. Alismail, H. Fattahi, J. Brons, O. Pronin, H. G. Barros, Z. Major, T. Metzger, D. Sutter, and F. Krausz, Opt. Lett. **42**, 1381 (2017).
2. H. Höppner, A. Hage, T. Tanikawa, M. Schulz, R. Riedel, U. Teubner, M. J. Prandolini, B. Faatz, and F. Tavella, New J. Phys. **17**, (2015).
3. T. Eidam, S. Hanf, E. Seise, T. V Andersen, T. Gabler, C. Wirth, T. Schreiber, J. Limpert, and A. Tünnermann, Opt. Lett. **35**, 94–96 (2010).
4. S. Piehler, B. Weichelt, A. Voss, M. A. Ahmed, and T. Graf, Opt. Lett. **37**, 5033 (2012).
5. T. Eidam, C. Wirth, C. Jauregui, F. Stutzki, F. Jansen, H.-J. Otto, O. Schmidt, T. Schreiber, J. Limpert, and A. Tünnermann, Opt. Express **19**, 13218–13224 (2011).
6. T. Y. Fan, IEEE J. Sel. Top. QUANTUM Electron. **11**, 567 (2005).
7. J. Limpert, A. Klenke, M. Kienel, S. Breitkopf, T. Eidam, S. Hadrich, C. Jauregui, and A. Tunnermann, IEEE J. Sel. Top. Quantum Electron. **20**, 268–277 (2014).
8. G. D. Goodno, C.-C. Shih, and J. E. Rothenberg, Opt. Express **18**, 25403–25414 (2010).
9. J. Le Dortz, A. Heilmann, M. Antier, J. Bourderionnet, C. Larat, I. Fsaifes, L. Daniault, S. Bellanger, C. Simon Boisson, J.-C. Chanteloup, E. Lallier, and A. Brignon, Opt. Lett. **42**, 1887 (2017).
10. M. Müller, M. Kienel, A. Klenke, T. Gottschall, E. Shestaev, M. Plötner, J. Limpert, and A. Tünnermann, Opt. Lett. **41**, 3439 (2016).
11. T. Zhou, Q. Du, T. Sano, R. Wilcox, and W. Leemans, Opt. Lett. **43**, 3269 (2018).
12. A. Vernaleken, J. Weitenberg, J. Sartorius, P. Russbueldt, W. Schneider, S. L. Stebbings, M. F. Kling, P. Hommelhoff, H.-D. Hoffmann, R. Poprawe, F. Krausz, T. W. Hänsch, and T. Udem, Opt. Lett. **36**, 3428 (2011).
13. V. V Lozovoy, I. Pastirk, and M. Dantus, Opt. Lett. **29**, 775 (2004).
14. T. M. Shay, V. Benham, J. T. Baker, A. D. Sanchez, D. Pilkington, and C. A. Lu, IEEE J. Sel. Top. Quantum Electron. **13**, 480–486 (2007).
15. S. B. Weiss, M. E. Weber, and G. D. Goodno, Opt. Lett. **37**, 455–7 (2012).
16. G. D. Goodno and S. B. Weiss, Opt. Express **20**, 14945 (2012).